# GRAIN DESTRUCTION IN INTERSTELLAR SHOCK WAVES

B.T. DRAINE

*Princeton University Observatory, Peyton Hall, Princeton, NJ 08544, USA;*
*draine@astro.princeton.edu*

**Abstract.** Interstellar shock waves can erode and destroy grains present in the shocked gas, primarily as the result of sputtering and grain-grain collisions. Uncertainties in current estimates of sputtering yields are reviewed. Results are presented for the simple case of sputtering of fast grains being stopped in cold gas. An upper limit is derived for sputtering of refractory grains in C-type MHD shocks: shock speeds $v_s \gtrsim 50 \, \mathrm{km \, s^{-1}}$ are required for return of more than 30% of the silicate to the gas phase. Sputtering can also be important for removing molecular ice mantles from grains in two-fluid MHD shock waves in molecular gas. Recent estimates of refractory grain lifetimes against destruction in shock waves are summarized, and the implications of these short lifetimes are discussed.

## 1. Introduction

A shock wave passing through interstellar gas may create a sufficiently hostile environment that grains present in the preshock gas may be appreciably altered or destroyed. This is important for several reasons:

1. Many elements (e.g., Ca, Fe, Si) are normally found to be substantially depleted from the interstellar gas – locked up in interstellar grains. Return of this material to the gas can alter the gas phase composition, possibly resulting in significant changes in the rate of radiative cooling, or in the strengths of diagnostic emission or absorption lines.
2. In very fast ($v_s \gtrsim 500 \, \mathrm{km \, s^{-1}}$) shocks, collisions of electrons with the grains may dominate the cooling of the gas until the grains are substantially destroyed (Ostriker & Silk 1973; Draine 1981). Detailed understanding of the cooling of the shocked gas then hinges upon understanding what happens to the grains.
3. The cool dense gas behind a radiative shock wave may provide conditions suitable for molecule formation, but this will depend on whether grains survive to catalyze the formation of $H_2$. Grain catalysis of $H_2$ can be important as a source of heat (the energy released in $H_2$ formation), or as a source of molecules which may be significant coolants or useful diagnostics.
4. The grains may be dynamically important: charged grains may dominate the coupling of magnetic fields to the gas when the fractional ionization is low, and fast grains may even account for a significant fraction of the energy and momentum flux under some conditions (McKee et al. 1987).
5. Finally, shock-processing plays a major role in the overall evolution of interstellar dust! Our best estimates of the effects of interstellar shocks point to refractory grain lifetimes considerably shorter than the time scale for recycling the interstellar medium through stars; these estimates are difficult to reconcile with the observed extreme depletions of elements such as Fe and Ca.

Some contemporary views on the nature of the interstellar grain population are described in §2. Sputtering is thought to be the dominant grain destruction mechanism; current estimates for sputtering yields are reviewed in §3. The sputtering of grains being decelerated in cold gas is considered in §4, and in §5 the fate of grains in realistic shock models is discussed, including the sputtering of grains in two-fluid C-type shocks. The implications for our understanding of interstellar grain evolution are discussed in §6; a brief summary is provided in §7.

## 2. What are Interstellar Grains?

Any discussion of the effects of interstellar shocks on grains must begin with some assumptions about the properties of these grains – their size distribution, structure, composition, and overall numbers. Unfortunately, we remain quite uncertain on many of these points. Current grain models are discussed in several recent reviews (Mathis 1993; Draine 1995).

Models for interstellar grains are constrained to reproduce the observed extinction, scattering, polarization of starlight, and infrared emission, while not consuming elements in excess of what is allowed by cosmic abundances (minus the abundances observed to be present in the gas). These constraints are not sufficient to uniquely determine the model, and a number of different grain models are currently under consideration.

All models agree that:
- Approximately $^2/_3$ of the grain mass is in sizes $a > 0.05 \mu$m – these are required to explain the observed extinction, polarization, and scattering at visible wavelengths.
- $\sim 20\%$ of the grain mass is in very small grains with $a < 0.025 \mu$m, these being required to explain the steep rise in ultraviolet extinction.
- Most of the interstellar Si is locked up in silicate grains – these are required to explain the strong 10 and $18\mu$m absorption/emission features.
- Most of the available Mg, Si, Fe is in grains, as well as a substantial ($^1/_3$ – $^2/_3$?) fraction of the C, consistent with observed interstellar gas-phase abundances – see Figure 1.

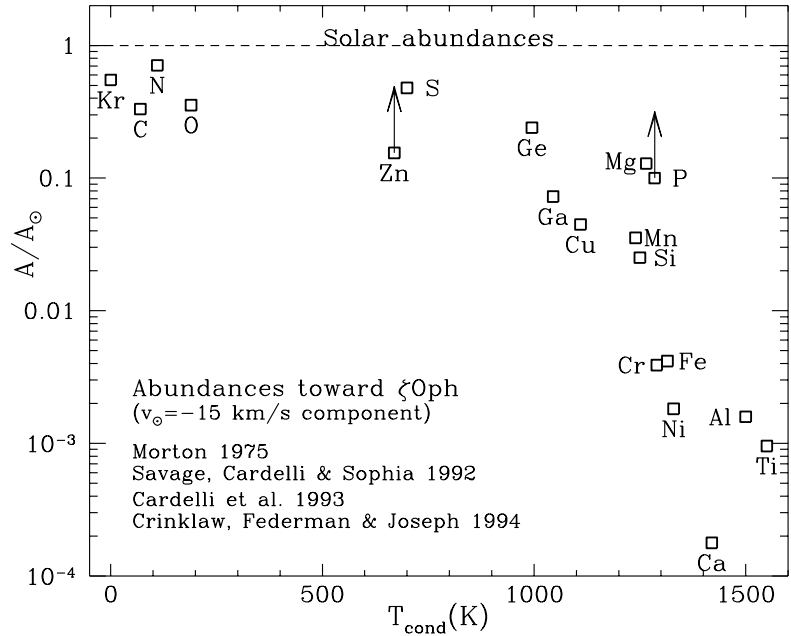

Fig. 1. Gas-phase abundances, relative to solar, toward ζOph, plotted against "condensation temperature" (see Field 1974).

There are substantial differences among the different grain models, some of which are as follows:

1. A mixture of bare graphite and and bare silicate grains, with approximately power-law size distributions extending from $.005 < a < 0.25\mu$m (Mathis, Rumpl, & Nordsieck 1977; Draine & Lee 1984), supplemented by a population of ultrasmall grains (Draine & Anderson 1985) including polycyclic aromatic hydrocarbon (PAH) molecules (Léger and Puget 1984).

2. Silicate cores with a carbonaceous coating, plus a poulation of small bare graphite grains, plus PAH molecules (Désert, Boulanger, & Puget 1990).

3. Silicate cores with hydrogenated amorphous carbon coatings (Duley, Jones & Williams 1989).

4. Small pieces of graphite, silicate, and amorphous carbon – some as individual small grains, but mostly coagulated into mixed fluffy structures (Mathis & Whiffen 1989).

## 3. Grain Destruction Mechanisms

3.1. SPUTTERING

Sputtering appears to be the dominant grain destruction mechanism in the galaxy. Sputtering rates depend on the sputtering yield $Y(E,\theta)$, the average number of sputtered atoms (or molecules) per incident atom or ion. Refractory materials are believed to sputter one atom at a time; molecular ices (e.g., $H_2O$) will sputter molecule by molecule. Sputtering yields depend on the projectile type, energy, and incidence angle, and on the target composition. Experimental sputtering data have been reviewed by Andersen & Bay (1981) and Betz & Wehner (1983). Yields for H and He incident on candidate grain materials have been estimated by Barlow (1978a), Draine & Salpeter (1979a, hereafter DS79), and Tielens et al. (1994, hereafter TMSH94).

The threshold energy $E_T$ is a critical parameter. Barlow found that $E_T \approx 4U_0$ for a wide range of projectile-target combinations; DS79 extended this to

$$E_T = \max\left(4U_0, \frac{(m_P + m_T)^2}{4 m_P m_T} U_0\right) \qquad (1)$$

to allow for inefficient energy transfer for large or small values of the projectile/target mass ratio $m_P/m_T$.

For $E > E_T$, Barlow took the sputtering yield to be a linear function of energy; DS79 and TMSH94 adopted more complicated fits, allowing for the decrease in sputtering yields at high energies, based upon the theory of Sigmund (1969; 1981). We recommend using either the DS79 or TMSH94 yields.

Estimated sputtering yields for H and He incident on solid carbon are shown in Figure 2. The estimate of DS79 uses the binding energy $U_0 = 7.37$eV appropriate for graphite; TMSH94 have noted that the surface of a graphite grain undergoing sputtering will be damaged, and adopt an effective binding energy $U_0 = 4$eV. The differences at low energy between the DS79 and TMSH94 estimates for carbon arise from this difference in adopted binding energy. No reliable data appears to be available for silicates; as a similar material, we show in Figure 3 theoretical and experimental results for $SiO_2$. In this case the DS79 and TMSH94 estimates are quite similar. Figure 4 shows predicted yields for $MgFeSiO_4$, with an average binding energy per atom $U_0 = 5.65$eV. Once again, the DS79 and TMSH94 estimates are very similar.

The above yields are for normal-incidence monoenergetic particles, and it is necessary to average over incidence angle and energy. For a neutral spherical target moving with drift velocity $v_d$, the rate of sputtering due to a projectile species with density $n$ and mass $m$ is (DS79)

$$\frac{dN}{dt} = \pi a^2 n \left(\frac{8kT}{\pi m}\right)^{1/2} \int_0^\infty dx\, x^2 e^{-x^2} \sinh(2sx) \langle Y(E = x^2 kT) \rangle_\theta \quad (2)$$

$$s^2 \equiv m v_d^2 / 2kT \quad (3)$$

where the angle-averaged sputtering yield

$$\langle Y(E) \rangle_\theta \equiv 2 \int_0^{\pi/2} Y(E, \theta) \sin\theta \cos\theta\, d\theta \approx 2Y(E, \theta = 0) \quad . \quad (4)$$

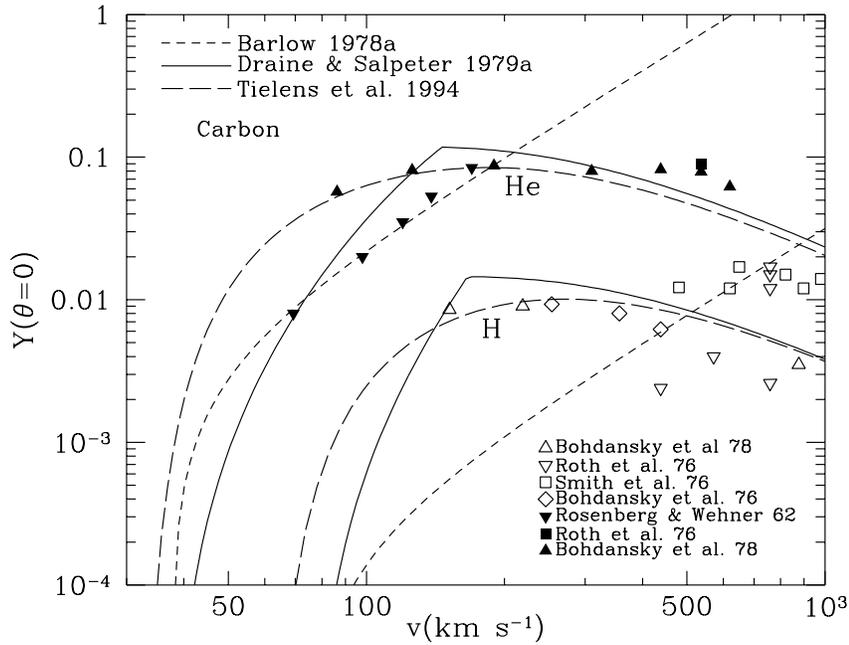

Fig. 2. Normal-incidence sputtering yield for H and He on solid carbon, vs. projectile velocity. Curves are theoretical estimates; symbols are experimental measurements.

Sputtering rates should be summed over all important projectile species (H and He generally dominate for cosmic abundances). The effects of grain charging are readily included (DS79).

3.2. GRAIN-GRAIN COLLISIONS

Grain-grain collisions at relative speeds exceeding a few km/s can destroy grains. The physics of grain-grain collisions has recently been discussed by TMSH94. For equal-sized refractory grains colliding at $v_s = 20\,\mathrm{km\,s^{-1}}$, they estimate $\sim 50\%$ vaporization.

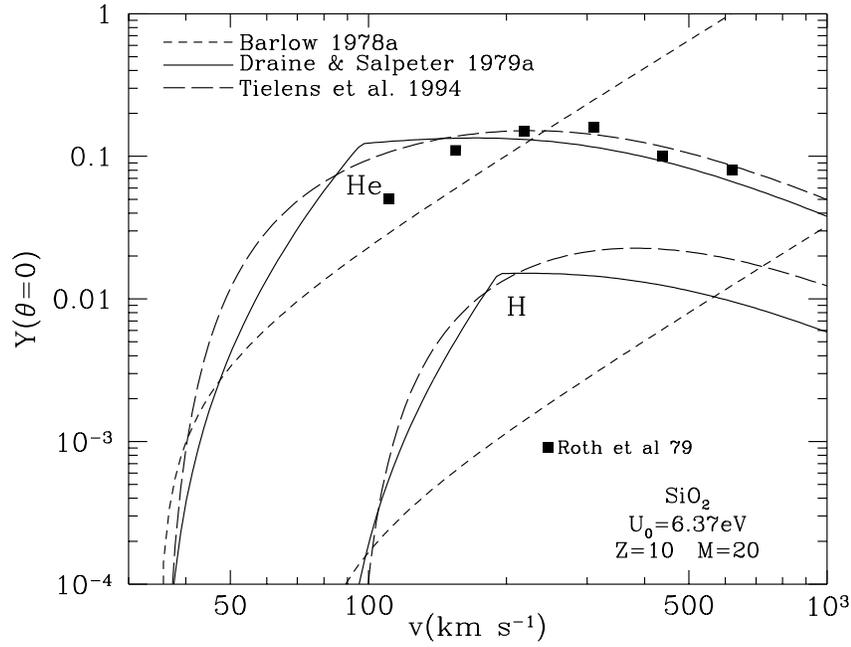

Fig. 3. Normal-incidence sputtering yields for H and He on $SiO_2$. Experimental data of Roth, Bohdansky, & Ottenberger (1979) is taken from Betz & Wehner (1983).

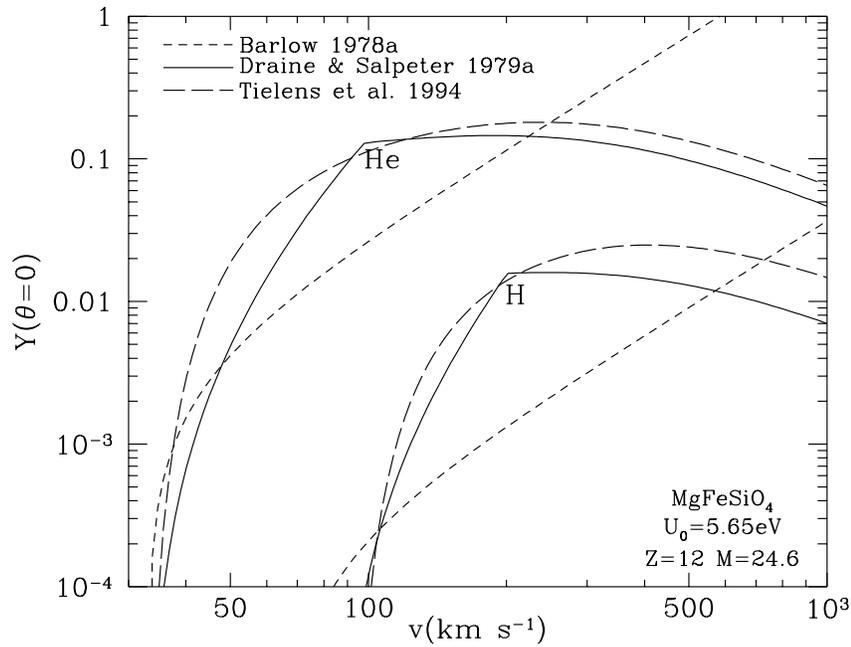

Fig. 4. Normal-incidence sputtering yields for H and He on $MgFeSiO_4$.

### 3.3. CHEMISPUTTERING

Carbon grains may be subject to chemical attack by impinging H and O atoms. Barlow & Silk (1977) and Barlow (1978b) concluded that such chemisputtering would rapidly destroy carbon grains in HI or HII regions. However, it appears that *removal* of carbon atoms by impinging H and O is thermally-activated, and is very inefficient at the low temperatures of interstellar grains (Draine 1979), so this does not appear to be an important destruction mechanism. Impinging hydrogen may, however, hydrogenate the carbon (cf. Furton & Witt 1993).

## 4. Sputtering of Hypersonic Grains

Sputtering of grains in single-fluid shocks has been discussed by many authors, including Shull (1977,1978), Cowie (1978), Barlow (1978a), Draine & Salpeter (1979b), Seab & Shull (1983), McKee et al. (1987), and Jones et al. (1994). Following Cowie (1978), it is instructive to consider a grain with an initial velocity $v_0$ in cold gas. It will decelerate by gas drag, and, if moving rapidly enough, will be subject to sputtering. If the motion is hypersonic, and He/H=0.1, then it is straightforward to show that the grain radius $a$ and velocity $v$ will vary as

$$4\pi \rho a^2 \frac{da}{dt} = -\pi a^2 v n_\mathrm{H} Y_\mathrm{eff}(v) m_T \tag{5}$$

$$\frac{4\pi}{3} \rho a^3 \frac{dv}{dt} = -\pi a^2 n_\mathrm{H} 1.4 m_\mathrm{H} v^2 \tag{6}$$

$$Y_\mathrm{eff}(v) \equiv \langle Y_\mathrm{H}(v) \rangle_\theta + 0.1 \langle Y_\mathrm{He}(v) \rangle_\theta \tag{7}$$

where $\langle Y_\mathrm{H}(v) \rangle_\theta$ and $\langle Y_\mathrm{He}(v) \rangle_\theta$ are the angle-averaged yields for H and He with velocity $v$. Dividing (5) by (6) we obtain

$$d\ln a^3 = \left(\frac{m_T}{1.4 m_\mathrm{H}}\right) Y_\mathrm{eff}(v) d\ln v \tag{8}$$

This equation has been integrated numerically to compute the fractional volume sputtered in slowing a grain down from some initial velocity $v_0$; the results are shown in Fig. 5. It should be recognized that this is neither an upper nor a lower bound to the sputtering in an actual shock. Charged grains may be decelerated by plasma drag in addition to direct collisions, reducing the sputtering. On the other hand, thermal motions of the atoms and ions will tend to increase sputtering rates, and betatron acceleration can accelerate the grains (see below).

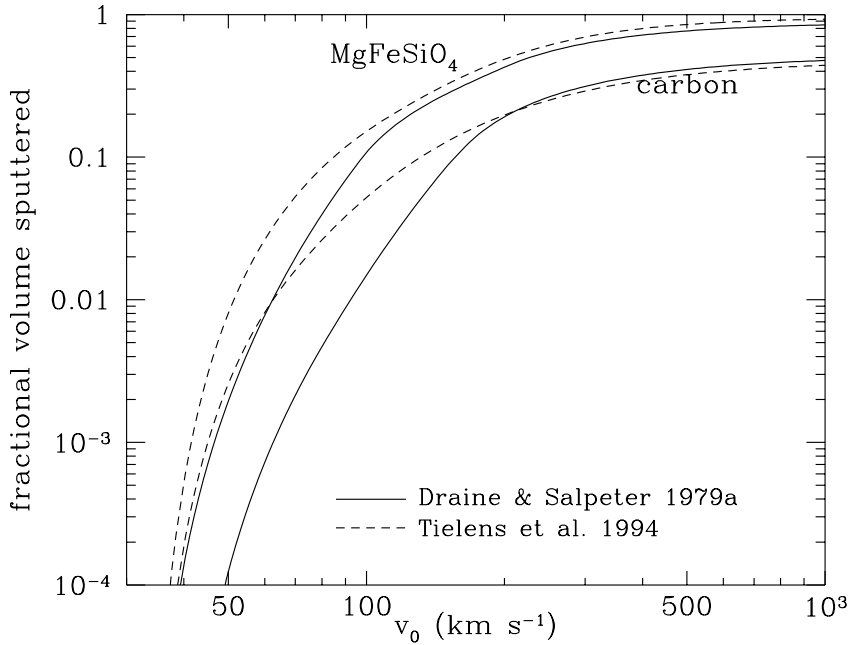

Fig. 5. Fractional volume sputtered from a grain, with initial velocity $v_0$, when stopped by cold gas.

## 5. Shock Structure and Grain Dynamics

The structure of interstellar shocks is complicated by many issues, including collisionless processes, magnetic stresses, the possible acceleration of cosmic rays to dynamically significant pressures, multifluid behavior, and shock instabilities; these have been reviewed recently elsewhere (Draine & McKee 1993). Here we confine discussion to two simple cases.

### 5.1. Fast Single-Fluid Shocks

When the fractional ionization is sufficiently high, ions, neutrals, and magnetic field are well-coupled and flow together as a single fluid – this is the case in shocks in ionized gas and in HI clouds. In a strong shock, the gas is compressed by a factor of 4 at the jump front. In the "shock frame", the gas is decelerated to $v_s/4$, but the grains cross the jump front with their preshock velocity $v_s$, and hence have a velocity $0.75v_s$ relative to the gas. If a transverse magnetic field is present, the (charged) grains will begin to orbit around field lines with an orbital velocity $0.75v_s$. Calculations of grain erosion by sputtering must take into account these substantial gas-grain velocities, particularly with regard to sputtering by helium. Gas drag will tend to reduce the orbital velocity, but if the gas cools and the magnetic field lines are compressed on a time short compared to the time scale for gas drag to slow the grain (but on a time scale long compared to the grain gyroperiod), then Spitzer (1977) pointed out that the grains can actually be *accelerated* by the "betatron" mechanism.

Detailed calculations of grain destruction in shocks, including betatron acceleration, have been carried out (Draine & Salpeter 1979b; Seab & Shull 1983; McKee et al. 1987; Jones et al. 1994); these studies generally find substantial ($> 30\%$) sputtering of refractory grains in radiative shocks with $v_s \gtrsim 200\,\mathrm{km\,s^{-1}}$. The most recent work by Jones et al. finds sputtering to be more important than grain-grain collisions for $v_s \gtrsim 60\,\mathrm{km\,s^{-1}}$, with grain-grain collisions destroying a few percent of the grain volume at $v_s = 50\,\mathrm{km\,s^{-1}}$.

## 5.2. C-Type Shocks in Molecular Gas

When the fractional ionization is sufficiently low, shocks in magnetized gas develop a two-fluid structure, since the charged particles are coupled to the magnetic field, but the neutral particles "know" about the magnetic field only via infrequent collisions with the ions. The resulting shock is often "C-type" (Draine 1980), with large streaming velocities between the ions and the neutrals. Because the dust grains are charged, they are strongly affected by the magnetic field, which causes the dust grains to move through the neutral gas with a drift velocity which can approach the ion-neutral streaming velocity. This drift through the neutrals can lead to sputtering of molecular ices (Draine, Roberge & Dalgarno 1983; Flower & Pineau des Forêts 1994)) and possibly even refractory grains (Flower & Pineau des Forêts 1995). Flower, Pineau des Forêts, & Walmsley (1995) have proposed that hot $NH_3$ observed toward Sgr B2 was injected into the gas by sputtering of ice mantles in a $v_s \approx 25\,\mathrm{km\,s^{-1}}$ shock.

In these C-type shocks the gas remains relatively cold, and sputtering will be determined largely by the motion of the grain through the gas. We have seen above the sputtering which occurs when a grain is stopped in cold gas (see Fig. 5); we see that even for an initial gas-grain velocity of $50\,\mathrm{km\,s^{-1}}$ only $\sim 0.2\%$ of a silicate grain will be eroded (using the DS79 sputtering yields). However, in an MHD shock the grains will be driven through the gas by the magnetic field. What happens here is that in the "shock frame" the grains are rapidly decelerated by magnetic forces; the neutral gas continues to stream until it is brought to rest by collisions with ions or charged grains. If the ionization is low, then collisions with grains may dominate, and each grain has to collide with many times its own mass of gas, leading to greater sputtering than would occur in the absence of magnetic field (in which case the results shown in Fig. 5 would apply).

For molecular ices, models of MHD shocks in dense molecular gas lead to erosion of $\sim 10\%$ of water ice mantles at shock speeds $v_s \approx 20\,\mathrm{km\,s^{-1}}$ (Draine, Roberge, & Dalgarno 1983). Flower & Pineau des Forêts (1994) found that shock speeds as small as $10\,\mathrm{km\,s^{-1}}$ were sufficient to fully erode ice mantles, but they assumed a sputtering threshold $E_T = 0.52\mathrm{eV}$; for $\mathrm{H_2O}$ ice, with $U_0 = 0.53\mathrm{eV}$, the sputtering threshold should be about 4 times larger [see eq. (1)] than the value adopted by Flower & Pineau des Forêts.

We may obtain an upper limit to magnetic-field-driven sputtering in C-type shocks by a fairly simple argument. Suppose we have grains of a single radius $a$. The grain erosion rate is given by

$$4\pi a^2 v_g \frac{da}{dx} = -\pi a^2 v_{\mathrm{rel}} n_\mathrm{H} Y_{\mathrm{eff}}(v_{\mathrm{rel}}) m_T \tag{9}$$

where $v_g$ is the grain velocity and $v_{\mathrm{rel}} = v_n - v_g$ is the streaming velocity between the neutrals and the grains. In the shock frame, the momentum conservation equation for the cold neutral gas is

$$\frac{d}{dx}(\rho_n v_n^2) = F_{ni} + F_{ng} \tag{10}$$

where $\rho_n$ and $v_n$ are the density and velocity of the neutral gas, and $F_{ni}$ and $F_{ng}$ are the force per volume on the neutral gas due to collisions with streaming ions and dust grains, respectively. We have

$$F_{ng} = n_g \pi a^2 \rho_n v_{\mathrm{rel}}(v_g - v_n) \tag{11}$$

If we define

$$\alpha \equiv \frac{F_{ng}}{F_{ng} + F_{ni}} \tag{12}$$

we may divide (9) by (10) to obtain

$$d(a/a_i)^3 = R \frac{m_T}{1.4 m_\mathrm{H}} Y_{\mathrm{eff}}(v_{\mathrm{rel}}) d\ln v_n \tag{13}$$

$$R \equiv \alpha \frac{v_n}{v_n - v_g} \frac{\rho_n v_n}{(4\pi/3)\rho a_i^3 n_g v_g} \tag{14}$$

where $a_i$ is the initial grain radius. If we now assume $v_g \ll v_n$ in the region of the flow where sputtering is important, then $v_{\mathrm{rel}} \approx v_n$. Since $\alpha < 1$, we have

$$R < \frac{\rho_n v_n}{(4\pi/3)\rho a_i^3 n_g v_g} = \frac{\text{initial gas mass}}{\text{initial grain mass}} \approx 140 \tag{15}$$

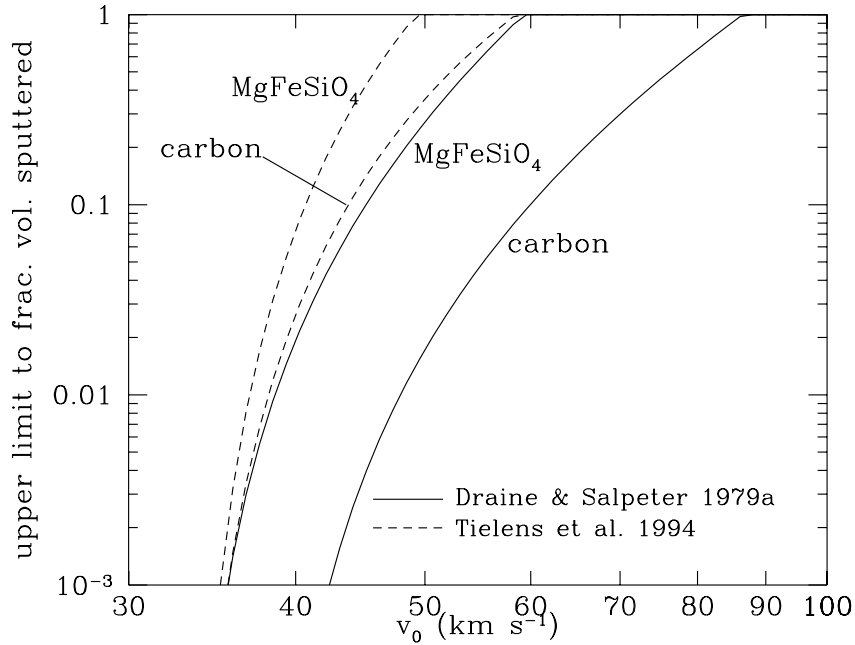

Fig. 6. Upper limit to volume sputtered in C-type MHD shock, using eq. (13) with $R < 140$ and $v_{\rm rel} < v_n$. Here $v_0 = v_s(1 - C^{-1})$ is the total velocity change in the shock, where $C$ is the overall compression ratio in the shock.

If we integrate eq.(13) with $R = 140$ and $v_{\rm rel} = v_n$ we obtain an upper limit on the the fractional volume sputtered in C-type shocks. Figure 6 shows the results of this calculation for carbon and silicate grains. It should be noted that this upper limit will be approached only in shocks where the fractional ionization remains very low, $n_e/n_{\rm H} < 10^{-6}$, since otherwise ion-neutral collisions are important, and $\alpha \ll 1$. Sputtering of refractory grains is negligible for $v_s < 35\,{\rm km\,s^{-1}}$.

## 6. Implications for Grain Evolution

The interstellar medium is a violent place, and shock waves are ubiquitous. Current estimates for frequencies of supernova-driven shocks, together with calculations of grain destruction in shocks, lead to lifetimes of $\sim 3 \times 10^8$ yrs for refractory grains (Draine & Salpeter 1979b; Jones et al. 1994). As discussed elsewhere (e.g., Draine 1990), this is a somewhat troubling result, since the timescale for replenishing the interstellar medium with "stardust" is considerably longer, of order $\sim 2 \times 10^9$ yr. As a result we would expect that only $\sim 10\%$ of the refractory elements (e.g., Si or Fe) should be locked up in "stardust". Since we observe 90% or more of some of these elements to be missing from the gas, we must conclude that they are locked up in solid material which was condensed *in the interstellar medium* rather than in stellar outflows. Since atoms and ions in the interstellar gas will collide with and may stick to preexisting grains, it is clear that a mechanism for such interstellar condensation exists – the problem is that the predicted *rates* for such condensation in diffuse clouds are simply too slow to explain the observed depletions. As a result, it seems to be necessary to postulate that most of this condensation takes place in dense molecular clouds; this, in turn, can only explain the depletions observed in diffuse clouds if there is quite rapid exchange of material between the diffuse and dense phases (Draine 1990).

In any case, the following conclusion appears inescapable: *interstellar grains are not (for the most part) stardust!*

## 7. Concluding Remarks

Shock waves can have substantial effects on interstellar grains, and in some cases the structure of the shock depends upon the grains mixed with the shocked gas.

Much of the physics involved in the modification of grains in shocked gas remains poorly understood. For example, sputtering yields, expecially at low energies, are quite uncertain – laboratory experiments with appropriate targets (e.g., silicates, amorphous carbon, graphite) and projectiles (H, $H_2$, He) are required for progress on this front.

Our modelling of shock waves in dusty gas leaves much room for progress. The grain trajectories are complicated by curving magnetic fields in oblique shocks, quantized changes in the grain charge, and grain-grain collisions.

Finally, the nature of the grains themselves is uncertain – in some models they are compact objects, in other models they are fluffy aggregates – and this lack of knowledge propagates into uncertainty on the fate of the grains in shock waves. Progress on this front will come slowly, as we gain a better understanding of interstellar grains from careful modelling of observed extinction, scattering, polarization, and emission.

This work was supported in part by NSF grant AST-9319283.This work was supported in part by NSF grant AST-9319283.